\documentclass[usenatbib]{mn2e}

\usepackage{epsf}

\newcommand\be{{\bmath e}}
\newcommand\bff{{\bmath f}}
\newcommand\bu{{\bmath u}}
\newcommand\bS{{\bmath S}}
\newcommand\bOmega{{\bmath\Omega}}
\newcommand\bnabla{{\bmath\nabla}}
\newcommand\real{\mathrm{Re}}
\newcommand\half{{\textstyle\frac{1}{2}}}
\newcommand\rmd{\mathrm{d}}
\newcommand\rme{\mathrm{e}}
\newcommand\rmf{\mathrm{f}}
\newcommand\rmi{\mathrm{i}}
\newcommand\rmv{\mathrm{v}}
\newcommand\f{\frac}
\newcommand\p{\partial}

\title[Tidal dissipation in rotating fluid bodies]
{Tidal dissipation in rotating fluid bodies: a simplified model}

\author[Gordon I. Ogilvie]{Gordon I. Ogilvie\\
Department of Applied Mathematics and Theoretical Physics,
University of Cambridge, Centre for Mathematical Sciences,\\
Wilberforce Road, Cambridge CB3 0WA
}

\begin{document}

\maketitle

\label{firstpage}
 
\begin{abstract}
We study the tidal forcing, propagation and dissipation of linear
inertial waves in a rotating fluid body.  The intentionally simplified
model involves a perfectly rigid core surrounded by a deep ocean
consisting of a homogeneous incompressible fluid.  Centrifugal effects
are neglected, but the Coriolis force is considered in full, and
dissipation occurs through viscous or frictional forces.  The
dissipation rate exhibits a complicated dependence on the
tidal frequency and generally increases with the size of the core.  In
certain intervals of frequency, efficient dissipation is found to
occur even for very small values of the coefficient of viscosity or
friction.  We discuss the results with reference to wave attractors,
critical latitudes and other features of the propagation of inertial
waves within the fluid, and comment on their relevance for tidal
dissipation in planets and stars.
\end{abstract}


\section{Introduction}
\label{s:introduction}

Tidal interactions determine the fate of short-period extrasolar
planets.  They affect their spin and orbital parameters, and in
extreme cases may lead to the destruction of planets as a result of
orbital decay or intense tidal heating.  Tidal evolution is also
important in close binary stars and in the satellite systems of the
planets of the solar system.

The efficiency of these processes is often parametrized by the tidal
quality factor~$Q$, which is a dimensionless inverse measure of the
dissipative properties of the body considered as a forced oscillator
\citep[e.g.][]{1966Icar....5..375G}.  The tidal~$Q$ of fluid bodies
such as stars and giant planets is difficult to calculate from first
principles, even in linear theory.  The tidal disturbance generally
consists of two parts.  One (the `equilibrium tide') is a
quasi-hydrostatic bulge that is carried around the body by a smooth
velocity field, while the other (the `dynamical tide') consists of
internal waves that are excited by the low-frequency tidal forcing and
may have a short wavelength.  Dissipation of the equilibrium tide can
occur through its interaction with turbulent convection, but the
damping rate is uncertain, especially when the tidal period is short
compared to the convective timescale
\citep{1966AnAp...29..489Z,1977A&A....57..383Z,1977ApJ...211..934G,1977Icar...30..301G,1997ApJ...486..403G,2007ApJ...655.1166P}.
This approach has perhaps been most successful in application to the
circularization of binaries containing a giant star
\citep{1995A&A...296..709V}, in which case the tidal period exceeds
the convective timescale.  To study dynamical tides one should
consider the excitation, propagation and dissipation of low-frequency
internal waves in rotating, stratified fluids.  Calculations have been
made for a variety of objects, using different simplifications and
approximations, by
\citet{1970A&A.....4..452Z,1975A&A....41..329Z,1977A&A....57..383Z},
\citet{1983MNRAS.203..581S,1984MNRAS.207..685S},
\citet{1995MNRAS.277..471S}, \citet{1997MNRAS.291..633S},
\citet{1997MNRAS.291..651P}, \citet{1997ApJ...484..866L},
\citet{1998ApJ...502..788T}, \citet{1998ApJ...507..938G},
\citet{1999A&A...341..842W}, \citet{2002A&A...386..211S},
\citet{2004ApJ...610..477O,2007ApJ...661.1180O},
\citet{2005ApJ...635..674W,2005ApJ...635..688W},
\citet{2005MNRAS.364L..66P} and \citet{2007MNRAS.376..682I}.  This
approach has perhaps been most successful for early-type stars, in
which gravity (or inertia-gravity) waves are excited near the base of
the radiative envelope and propagate outwards until they are
dissipated by radiative damping; in this case estimates can be made
rather simply for the dissipation rate (or tidal torque) that are
independent of the details of the wave damping mechanism
\citep{1989ApJ...342.1079G}.

In recent work on dynamical tides in rotating giant planets and stars
\citep{2004ApJ...610..477O,2007ApJ...661.1180O} we have emphasized the
role of inertial waves in convective regions, which are nearly
adiabatically stratified and do not support gravity waves \citep[see
also][]{2005ApJ...635..674W,2005ApJ...635..688W,2005MNRAS.364L..66P,2007MNRAS.376..682I}.
Inertial waves \citep{1968Greenspan} propagate in a uniformly rotating
fluid at frequencies $\omega$ smaller in magnitude than twice the spin
frequency $\Omega$, as seen in a frame rotating with the fluid.  The
group velocity of an inertial wave is proportional to its wavelength
and is inclined at an angle $\lambda=\arcsin|\omega/2\Omega|$ to the
rotation axis.  The behaviour of rays propagating within a spherical
annulus (a thick spherical shell) is very complicated and sensitive to
the value of $\lambda$ \citep{2001JFM...435..103R}.  It generally
involves the focusing of rays towards limit cycles known as wave
attractors.  Another important feature is the existence of a critical
latitude (equal to $\lambda$) at which the rays are tangent to the
core and a singularity is introduced into solutions of the inviscid
wave equations.  Numerical investigations indicate that the tidally
forced disturbances in this frequency range are concentrated in narrow
beams whose width diminishes as the viscosity $\nu$ is reduced
\citep{2004ApJ...610..477O}.  In certain intervals of $\omega/\Omega$,
a dissipation rate that appears to be asymptotically independent of
$\nu$ may be achieved.  However, the dissipation rate varies in a
complicated way with the tidal frequency; presumably this occurs
because of the sensitivity of the ray propagation to the value of
$\lambda$, and depends on the waves being reflected from the inner and
outer boundaries.

In an attempt to understand aspects of this behaviour, we have
investigated a variety of reduced models.  Using a prototypical
partial differential equation for internal waves
\citep{2005JFM...543...19O}, we studied the response to periodic
forcing in a two-dimensional domain in which the rays are focused
towards a single wave attractor, as occurs in the experiments of
\citet{1997Natur.388..557M} and \citet{2003JFM...493....59M}.  We
constructed an asymptotic analytical solution and showed that a
non-zero dissipation rate is achieved in the astrophysically relevant
limit of small viscosity (i.e.\ small Ekman number).  This limiting
dissipation rate is independent of the magnitude and even of the form
of the small-scale damping mechanism for the waves, which is very
promising for astrophysical applications.  We confirmed the analytical
solution in detail using numerical methods.

The method used by \citet{2005JFM...543...19O} cannot be applied
directly to the problem of tidal forcing when the fluid occupies a
spherical annulus.  One technical complication is that the hyperbolic
equation describing the structure of inviscid inertial waves in the
meridional plane of a spherical system does not possess Riemann
invariants, which makes it difficult to describe the propagation of
the solution along the characteristics (which are also the ray paths
described above).  More importantly, the spherical annulus has a
concave inner boundary with a critical latitude singularity.  Although
such a feature could be introduced into the problem considered by
\citet{2005JFM...543...19O} by choosing a domain of an appropriate
shape, separate consideration would have to be given to the waves
generated by the singularity.

In the present paper we consider a different idealized model that
displays the essential features of inertial waves propagating within a
spherical annulus and allows a detailed numerical investigation.  It
involves a perfectly rigid core surrounded by a deep ocean consisting
of a homogeneous incompressible fluid.  Centrifugal effects are
neglected, but the Coriolis force is considered in full, and
dissipation occurs through viscous or frictional forces.  Although the
assumption of a homogeneous incompressible fluid is made for maximal
simplicity, the model may be more or less directly applicable to
planets or moons involving a deep ocean.  We formulate the problem of
tidal forcing in Section~\ref{s:tidal}, introduce a further
simplification in Section~\ref{s:forced} and present numerical
solutions in Section~\ref{s:numerical}.  The results are discussed in
Section~\ref{s:discussion} and a conclusion follows in
Section~\ref{s:conclusion}.

\section{The tidal problem}
\label{s:tidal}

\subsection{Basic equations}

We consider a simplified model of a planet in uniform rotation with
angular velocity $\bOmega=\Omega\,\be_z$.  Let $(r,\theta,\phi)$ be spherical
polar coordinates in the rotating frame.  A perfectly rigid core
occupies the region $0<r<\alpha R$ and is surrounded by a homogeneous
incompressible fluid with a free surface at $r=R$.  We neglect the
centrifugal distortion, which is of second order in $\Omega$.  Let the
mean density of the system be $\bar\rho$ and that of the fluid
$\rho=\beta\bar\rho$.  The surface gravity is then
\begin{equation}
  g=\f{4}{3}\pi G\bar\rho R.
\end{equation}

We consider the long-term response of the system to a tidal
gravitational potential $\real[\Psi(r,\theta,\phi)\,\rme^{-\rmi\omega
t}]$, where $\omega$ is the (real) tidal frequency in the rotating
frame.  The linearized equations within the fluid are
\begin{equation}
  -\rmi\omega\bu+2\bOmega\times\bu=-\bnabla W+\bff,
\end{equation}
\begin{equation}
  W=\f{p'}{\rho}+\Phi'+\Psi,
\end{equation}
\begin{equation}
  \bnabla\cdot\bu=0,
\end{equation}
\begin{equation}
  \nabla^2\Phi'=4\pi G\rho'=-4\pi G\rho\f{u_r}{\rmi\omega}\delta(r-R),
\end{equation}
where $\bu$, $\rho'$, $p'$ and $\Phi'$ are the Eulerian perturbations
of velocity, density, pressure and gravitational potential, assumed to
have the same form of time-dependence as the tidal potential, and
$\delta$ is the Dirac delta function.  We include a dissipative force
(per unit mass) $\bff$, in the form of either a Navier--Stokes viscous
force
\begin{equation}
  \bff_\rmv=\nu\nabla^2\bu,
\end{equation}
where $\nu$ is the kinematic viscosity, or a frictional damping force
\begin{equation}
  \bff_\rmf=-\gamma\bu,
\end{equation}
where $\gamma$ is the frictional damping coefficient.  The latter case is
artificial but is mathematically and computationally more convenient.
In \citet{2005JFM...543...19O} we found that the asymptotic
dissipation rate in a wave attractor is identical for viscous and
frictional forces.  We set $\bff=\bff_\rmv+\bff_\rmf$ in this section,
although in the numerical analysis below we consider each type of
dissipative force separately.

The time-averaged energy dissipation rate is
\begin{equation}
  D=-\f{1}{2}\,\real\int_V\rho\bu^*\cdot\bff\,\rmd V=-\f{1}{2}\,\real\int_S\rho W\bu^*\cdot\rmd\bS,
\end{equation}
where $S$ is the bounding surface of the volume $V$ occupied by the
fluid.  The last equality follows from the linearized equations and
the divergence theorem.  For viscous and frictional damping we have
\begin{equation}
  D_\rmv=\int_V\rho\nu e_{ij}^*e_{ij}\,\rmd V,
\end{equation}
\begin{equation}
  D_\rmf=\f{1}{2}\int_V\rho\gamma|\bu|^2\,\rmd V,
\end{equation}
where $e_{ij}=\half(\p_iu_j+\p_ju_i)$ is the rate-of-strain tensor in
Cartesian coordinates.  In the viscous case, either no-slip or
stress-free boundary conditions are required to ensure that there is
no viscous flux of energy out of the fluid volume.

\subsection{Projection on to spherical harmonics}

To solve the linearized equations, we introduce a spheroidal--toroidal
decomposition and a projection on to spherical harmonics
\citep{1966AnAp...29..313Z}.  Thus
\begin{equation}
  u_r=\sum a_n(r)Y_n^m(\theta,\phi),
\end{equation}
\begin{equation}
  u_\theta=r\sum\left[b_n(r)\f{\p}{\p\theta}+\f{c_n(r)}{\sin\theta}\f{\p}{\p\phi}\right]Y_n^m(\theta,\phi),
\end{equation}
\begin{equation}
  u_\phi=r\sum\left[\f{b_n(r)}{\sin\theta}\f{\p}{\p\phi}-c_n(r)\f{\p}{\p\theta}\right]Y_n^m(\theta,\phi),
\end{equation}
\begin{equation}
  W=\sum W_n(r)Y_n^m(\theta,\phi),
\end{equation}
\begin{equation}
  \Phi'=\sum\Phi'_n(r)Y_n^m(\theta,\phi),
\end{equation}
where the sums are carried out over integers $n\ge m\ge0$, and
\begin{equation}
  Y_n^m(\theta,\phi)=\left[\f{(2n+1)(n-m)!}{4\pi(n+m)!}\right]^{1/2}P_n^m(\cos\theta)\,\rme^{\rmi m\phi}
\end{equation}
is a spherical harmonic, satisfying the orthonormality relation
\begin{equation}
  \int_0^{2\pi}\!\int_0^\pi[Y_n^m(\theta,\phi)]^*Y_{n'}^{m'}(\theta,\phi)\sin\theta\,\rmd\theta\,\rmd\phi=\delta_{mm'}\,\delta_{nn'}.
\end{equation}
Bearing in mind the linearity of the problem, we consider a forcing
potential in the form of a single solid spherical harmonic,
\begin{equation}
  \Psi=\Psi_\ell(r)Y_\ell^m(\theta,\phi),
\end{equation}
for some $\ell\ge m$.  (Although $\Psi_\ell\propto r^\ell$, only the
value $\Psi_\ell(R)$ of the potential at the free surface is
significant in the case of an incompressible fluid.)  The dominant
tidal potentials are the quadrupole components with $\ell=2$, and of
these the case $m=2$ is usually the most important.  The linearized
equations couple together different values of $n$ through the Coriolis
force, while different values of $m$ are decoupled.

Projecting the equations on to spherical harmonics, we obtain
\begin{eqnarray}
  \lefteqn{(-\rmi\omega+\gamma)a_n-2\rmi m\Omega rb_n}&\nonumber\\
  &&+2\Omega r[(n-1)q_nc_{n-1}-(n+2)q_{n+1}c_{n+1}]\nonumber\\
  &&=-\f{\rmd W_n}{\rmd r}-n(n+1)\f{\nu}{r^2}\left[a_n-\f{\rmd}{\rmd r}(r^2b_n)\right],
\label{aneq}
\end{eqnarray}
\begin{eqnarray}
  \lefteqn{-\rmi\omega_nr^2b_n-\f{2\rmi m\Omega ra_n}{n(n+1)}}&\nonumber\\
  &&+2\Omega r^2[(n-1)\tilde q_nc_{n-1}+(n+2)\tilde q_{n+1}c_{n+1}]=-W_n\nonumber\\
  &&+\nu\left[\f{2a_n}{r}+\f{1}{r^2}\f{\rmd}{\rmd r}\left(r^4\f{\rmd b_n}{\rmd r}\right)-(n-1)(n+2)b_n\right],
\label{bneq}
\end{eqnarray}
\begin{eqnarray}
  \lefteqn{-\rmi\omega_nr^2c_n+2\Omega r(\tilde q_na_{n-1}-\tilde q_{n+1}a_{n+1})}&\nonumber\\
  &&-2\Omega r^2[(n-1)\tilde q_nb_{n-1}+(n+2)\tilde q_{n+1}b_{n+1}]\nonumber\\
  &&=\nu\left[\f{1}{r^2}\f{\rmd}{\rmd r}\left(r^4\f{\rmd c_n}{\rmd r}\right)-(n-1)(n+2)c_n\right],
\label{cneq}
\end{eqnarray}
\begin{equation}
  \f{1}{r^2}\f{\rmd}{\rmd r}(r^2a_n)-n(n+1)b_n=0,
\label{dneq}
\end{equation}
where
\begin{equation}
  \omega_n=\omega+\f{2m\Omega}{n(n+1)}+\rmi\gamma,
\end{equation}
\begin{equation}
  q_n=\left(\f{n^2-m^2}{4n^2-1}\right)^{1/2},\qquad
  \tilde q_n=\f{q_n}{n}.
\end{equation}
These equations agree with those of \citet{1997JFM...341...77R} and
\citet{2004ApJ...610..477O} in the appropriate limits.  They are to be
solved for integers $n\ge m$, but in practice the system must be
truncated.

The potential perturbation satisfies
\begin{equation}
  \f{1}{r^2}\f{\rmd}{\rmd r}\left(r^2\f{\rmd\Phi_n'}{\rmd r}\right)-\f{n(n+1)}{r^2}{\Phi_n'}=-4\pi G\rho\f{a_n}{\rmi\omega}\delta(r-R),
\end{equation}
and the solution that is well behaved as $r\to0$ and as $r\to\infty$ is
\begin{equation}
  \Phi_n'=\f{4\pi G\rho R}{2n+1}\f{a_n(R)}{\rmi\omega}\times\cases{(r/R)^n,&$0<r<R$,\cr (r/R)^{-(n+1)},&$r>R$.}
\end{equation}

\subsection{Boundary conditions}

Since the inner boundary is impermeable, $a_n=0$ at $r=\alpha R$.  At
the outer boundary, the normal stress at the perturbed free surface
vanishes.  Thus $-p_n'-\rho g(a_n/\rmi\omega)+2\rho\nu(\rmd a_n/\rmd
r)=0$, i.e.
\begin{equation}
  W_n-2\nu\f{\rmd a_n}{\rmd r}+\left[1-\left(\f{3\beta}{2n+1}\right)\right]\f{g}{\rmi\omega}a_n=\Psi_\ell\delta_{n\ell}
\label{tidalbc}
\end{equation}
at $r=R$.  In the limits of low frequency and low viscosity,
$\omega^2\ll g\ell/R$ and $\nu\ll gR/\omega$, the first two terms are
negligible, and this equation amounts to specifying the radial
velocity at the surface.  Indeed, it amounts to saying that the radial
displacement of the surface is equal to the simple equilibrium tide
$-\Psi/g$, with a correction (the term involving $\beta$) due to the
self-gravity of the fluid.

When viscosity is included, additional boundary conditions are
required.  We assume that the tangential stresses vanish at the inner
and outer boundaries, so
\begin{equation}
  \f{a_n}{r^2}+\f{\rmd b_n}{\rmd r}=\f{\rmd c_n}{\rmd r}=0
\end{equation}
at $r=\alpha R$ and $r=R$.  The inner boundary could alternatively be
treated using the no-slip conditions $b_n=c_n=0$, which would result
in a prominent Ekman or oscillatory boundary layer.

\subsection{Dissipation}

The time-averaged energy equation for this system has the form
\begin{equation}
  \f{\rmd F}{\rmd r}=-d_\rmv-d_\rmf,
\end{equation}
where
\begin{eqnarray}
  \lefteqn{F=\f{1}{2}r^2\rho\,\real\sum_na_n^*\left(W_n-2\nu\f{\rmd a_n}{\rmd r}\right)}&\nonumber\\
  &&-n(n+1)\nu\left[b_n^*\left(a_n+r^2\f{\rmd b_n}{\rmd r}\right)+r^2c_n^*\f{\rmd c_n}{\rmd r}\right]
\end{eqnarray}
is the radial energy flux and
\begin{eqnarray}
  \lefteqn{d_\rmv=\f{1}{2}r^2\rho\nu\sum_nn(n+1)\left(\left|\f{a_n}{r}+r\f{\rmd b_n}{\rmd r}\right|^2+\left|r\f{\rmd c_n}{\rmd r}\right|^2\right)}\nonumber\\
  &&+3\left|\f{\rmd a_n}{\rmd r}\right|^2+(n-1)n(n+1)(n+2)(|b_n|^2+|c_n|^2),
\end{eqnarray}
\begin{equation}
  d_\rmf=\f{1}{2}r^2\rho\gamma\sum_n|a_n|^2+n(n+1)r^2(|b_n|^2+|c_n|^2)
\end{equation}
are the viscous and frictional dissipation rates per unit radius.  The
integrated energy equation, together with the boundary conditions,
relates the total dissipation rate to the surface forcing:
\begin{eqnarray}
  D&=&D_\rmv+D_\rmf\nonumber\\
  &=&\int_{\alpha R}^R(d_\rmv+d_\rmf)\,\rmd r\label{dint}\\
  &=&-\f{1}{2}\rho R^2\,\real\left[a_\ell(R)^*\Psi_\ell(R)\right].\label{dsurf1}
\end{eqnarray}

\subsection{Frictional non-rotating problem}
\label{fricnr}

In the absence of rotation the different spherical harmonics are
decoupled.  When only a frictional force is present, the solution
satisfying the inner boundary condition is
\begin{equation}
  a_\ell=A\left[\left(\f{r}{R}\right)^{\ell-1}-\alpha^{2\ell+1}\left(\f{R}{r}\right)^{\ell+2}\right],
\end{equation}
\begin{equation}
  b_\ell=\f{A}{r}\left[\f{1}{\ell}\left(\f{r}{R}\right)^{\ell-1}+\f{\alpha^{2\ell+1}}{\ell+1}\left(\f{R}{r}\right)^{\ell+2}\right],
\end{equation}
\begin{equation}
  W_\ell=(\rmi\omega-\gamma)Ar\left[\f{1}{\ell}\left(\f{r}{R}\right)^{\ell-1}+\f{\alpha^{2\ell+1}}{\ell+1}\left(\f{R}{r}\right)^{\ell+2}\right],
\end{equation}
where $A$ is a complex coefficient, and all other components vanish.
The outer boundary condition determines the value of $A$ according to
\begin{eqnarray}
  \lefteqn{(\rmi\omega-\gamma)AR\left(\f{1}{\ell}+\f{\alpha^{2\ell+1}}{\ell+1}\right)}&\nonumber\\
  &&+\left[1-\left(\f{3\beta}{2\ell+1}\right)\right]\f{g}{\rmi\omega}A\left(1-\alpha^{2\ell+1}\right)=\Psi_\ell(R),
\label{outerbc}
\end{eqnarray}
and the dissipation rate is
\begin{eqnarray}
  D&=&\f{1}{2}\rho\gamma|A|^2R^3\left(1-\alpha^{2\ell+1}\right)\left(\f{1}{\ell}+\f{\alpha^{2\ell+1}}{\ell+1}\right)\nonumber\\
  &=&-\f{1}{2}\rho R^2\left(1-\alpha^{2\ell+1}\right)\,\real[A^*\Psi_\ell(R)].
\end{eqnarray}

We specialize to the case of a quadrupolar tide ($\ell=2$).  The
dissipation rate can then be converted into a tidal quality factor $Q$
using the relation
\begin{equation}
  D=\f{15}{8Q'}\f{R|\omega||\Psi_2(R)|^2}{2\pi G}.
\end{equation}
(Here $Q'=3Q/2k_2$ is the modified tidal quality factor, $k_2$ being
the second-order potential Love number.  For a homogeneous
fluid body $k_2=3/2$ and $Q'=Q$.)  The solution of
equation~(\ref{outerbc}) is
\begin{equation}
  A=\f{\rmi\omega\Psi_2(R)}{g}B,
\label{aa}
\end{equation}
where $B$ is given by
\begin{equation}
  B^{-1}=(1-\alpha^5)(1-{\textstyle\f{3}{5}}\beta)-(1+{\textstyle\f{2}{3}}\alpha^5)\f{\omega(\omega+\rmi\gamma)}{2g/R}
\label{bb}
\end{equation}
and embodies various corrections (due to the
self-gravity of the fluid and the non-zero tidal frequency) to the
simple equilibrium tide.  Then
\begin{equation}
  \f{1}{Q'}=\f{|B|^2}{5}(1-\alpha^5)(1+{\textstyle\f{2}{3}}\alpha^5)\beta\f{\gamma|\omega|}{g/R}.
\end{equation}
Note that a resonance occurs with the $\ell=2$ surface gravity mode
(f~mode) as the tidal frequency is increased into the vicinity of the
dynamical frequency $(g/R)^{1/2}$; in the coreless case ($\alpha=0$,
$\beta=1$) this occurs at $\omega^2=4g/5R$.

\subsection{Viscous non-rotating problem}
\label{viscnr}

When the fluid is non-rotating and has only viscous dissipation,
the general solution of the linearized equations in the case $\ell=2$ is
\begin{eqnarray}
  \lefteqn{a_2=\f{6}{r^4}\left[A_1r^5+A_2+A_3(3-3\rmi kr-k^2r^2)\,\rme^{\rmi kr}\right.}&\nonumber\\
  &&\left.+A_4(3+3\rmi kr-k^2r^2)\,\rme^{-\rmi kr}\right],
\end{eqnarray}
\begin{eqnarray}
  \lefteqn{b_2=\f{1}{r^5}\left[3A_1r^5-2A_2-A_3(6-6\rmi kr-3k^2r^2+\rmi k^3r^3)\,\rme^{\rmi kr}\right.}&\nonumber\\
  &&\left.-A_4(6+6\rmi kr-3k^2r^2-\rmi k^3r^3)\,\rme^{-\rmi kr}\right],
\end{eqnarray}
\begin{equation}
  W_2=\f{\rmi\omega}{r^3}(3A_1r^5-2A_2),
\end{equation}
where $A_1$, $A_2$, $A_3$ and $A_4$ are complex coefficients and
\begin{equation}
  k=\left(\f{\rmi\omega}{\nu}\right)^{1/2}.
\end{equation}
Note that the terms involving an exponential correspond to viscous
boundary-layer solutions, while the other terms are solutions of the
inviscid problem \citep[cf.][]{1991GAFD...59..185R}.

With stress-free inner and outer boundary conditions, the solution for
the velocity in the limit of small $\nu$ is everywhere close to
the inviscid one.  The boundary-layer solutions, which decay rapidly
with distance, intervene near the boundaries to adjust the shear rate
to match the stress-free conditions.  Viscous dissipation is dominated
by the region away from the boundaries, for which the shear rate is
essentially that of the inviscid solution.  We then find
\begin{equation}
  D\approx\f{5}{6}(3+5\alpha^3-8\alpha^{10})\rho\nu R|A|^2,
\end{equation}
with $A$ as given by equations~(\ref{aa}) and~(\ref{bb}) (with
$\gamma=0$).  Thus
\begin{equation}
  \f{1}{Q'}\approx2|B|^2(1+{\textstyle\f{5}{3}}\alpha^3-{\textstyle\f{8}{3}}\alpha^{10})\beta\f{\nu|\omega|}{gR}.
\end{equation}

With a no-slip inner boundary condition, viscous dissipation would be
dominated by the shear layer on the inner boundary.  In this case the
dissipation rate would be approximately proportional to $\nu^{1/2}$,
rather than $\nu$, in the limit of small viscosity.

\subsection{Tidally forced inertial waves}

When rotation is included, the linearized equations support inertial
waves for frequencies in the range $-2<\omega/\Omega<2$.  This
can be seen by setting the dissipative force $\bff$ to zero and
deducing the Poincar\'e equation
\begin{equation}
  \omega^2\nabla^2W-(2\bOmega\cdot\bnabla)^2W=0
\end{equation}
for the modified pressure perturbation \citep{1968Greenspan}.  When
rendered in cylindrical or spherical polar coordinates for solutions
with the azimuthal dependence $\rme^{\rmi m\phi}$, this becomes a
second-order partial differential equation that is of hyperbolic type
for $0<\omega^2<4\Omega^2$.  The characteristics are straight lines in
the meridional plane inclined at an angle
$\lambda=\arcsin|\omega/2\Omega|$ to the rotation axis.

Although the Poincar\'e equation is homogeneous, the waves are forced
through the inhomogeneous outer boundary condition.  Since a
hyperbolic equation is generally ill posed when supplied with boundary
conditions on a closed surface, the inviscid problem generally has no
solution.  Either viscosity or friction formally removes the
hyperbolic character of the equation and leads to a well posed
problem.  The initial-value problem for an inviscid fluid is
also well posed.

\section{Radially forced oscillations}
\label{s:forced}

\subsection{Radial forcing}

We have seen that, in the limits of low frequency and low viscosity,
the outer boundary condition amounts to specifying the radial velocity
at the surface.  We therefore consider a slightly simpler problem in
which the free outer surface is replaced by a boundary (having zero
tangential viscous stress) on which the radial velocity $u_r$ is
prescribed and is in the form of a single spherical harmonic, i.e.
\begin{equation}
  a_n=U\delta_{n\ell}
\end{equation}
at $r=R$, where $U$ is an arbitrary (real) amplitude.  The dissipation
rate is then
\begin{equation}
  D=-\f{1}{2}\rho R^2U\,\real\left[W_\ell(R)\right].
\label{dsurf2}
\end{equation}

\subsection{Frictional problem}
\label{frictional}

When only a frictional force is present, we can eliminate variables in
favour of the radial velocity to obtain
\begin{eqnarray}
  \lefteqn{\left[\left(\f{n-1}{n}\right)\f{\tilde q_n^2}{\omega_{n-1}}+\left(\f{n+2}{n+1}\right)\f{\tilde q_{n+1}^2}{\omega_{n+1}}-\f{\omega_n}{4n(n+1)\Omega^2}\right]}&\nonumber\\
  &&\qquad\times(\p_y-n+1)(\p_y+n+2)a_n\nonumber\\
  &&+\f{\tilde q_n\tilde q_{n-1}}{\omega_{n-1}}(\p_y-n+1)(\p_y-n+3)a_{n-2}\nonumber\\
  &&+\f{\tilde q_{n+1}\tilde q_{n+2}}{\omega_{n+1}}(\p_y+n+2)(\p_y+n+4)a_{n+2}=0,
\label{aneq2}
\end{eqnarray}
where $y=\ln(r/R)$.

We truncate the system at spherical harmonic degree $L$ by setting
$a_n=0$ for $n>L$.  The truncated problem consists of a finite system
of homogeneous linear ordinary differential equations (ODEs) with
constant coefficients.  This possesses a basis of exponential
solutions with $a_n\propto\exp[(k-1)y]\propto r^{k-1}$ for suitable
values of $k$, in which the values of $a_n$ are related by a
three-term recurrence relation of the form
\begin{equation}
  X_na_{n-2}+Y_na_n+Z_na_{n+2}=0.
\end{equation}
In the case $\ell=m=2$, for example, the permitted velocity components
are $n=2,4,6,\dots,L$ and the coefficients of the recurrence relation
satisfy $X_2=Z_L=0$.  The system of ODEs is of order $L$ and its
exponential solutions are of the following form:
\begin{enumerate}
\item Solutions with $k=2,4,6,\dots,L$.  These can be normalized by
setting $a_2=1$, and iteration of the recurrence relation yields
$a_4,a_6,\dots,a_k$.  Beyond this point, $a_{k+2}=0$ because
$X_k=Y_k=0$.  (These are exact solutions of the untruncated problem.)
\item Solutions with $k=-L-1,-L+1,-L+3,\dots,-3$.  With the
normalization $a_L=1$, the recurrence relation yields
$a_{L-2},a_{L-4},\dots,a_{-k-1}$.  Then $a_{-k-3}=0$ because
$Y_{-k-1}=Z_{-k-1}=0$.  (These solutions are exact only in the
truncated problem.)
\end{enumerate}
In a related approach, \citet{1991GAFD...59..185R} obtained analytical
solutions of the viscous problem in terms of Bessel functions and
polynomials in $r$.

It is possible to solve the problem of forced inertial waves by
writing the solution as a linear combination of the exponential
solutions and applying the boundary conditions to determine the
unknown coefficients.  However, the solutions with large values of
$|k|$ vary by many orders of magnitude between the inner and outer
boundaries, except possibly in the case of a thin shell, and this
gives rise to a highly ill conditioned problem.  It would be
interesting to investigate whether this problem can nevertheless be
solved numerically for large values of $L$.  In this paper, however,
we mainly solve the ODEs using the Chebyshev collocation method, as
discussed in Section~\ref{s:method} below.

\subsection{Normal modes of the inviscid problem}
\label{s:modes}

Do there exist normal modes for inertial waves in a spherical annulus?
Since we are now considering a radially forced problem, normal modes
would be solutions of the same problem but with the homogeneous outer
boundary condition $a_n=0$.  These are free oscillations within rigid
boundaries, and should agree with the oscillation modes of a fluid
with a free outer surface in the limit of frequencies much lower than
those of surface gravity waves.  We consider a fluid without
dissipative forces; if it possesses a normal mode then the frictional
problem admits the same mode but with an additional damping rate of
$\gamma$.  (Viscous fluids certainly do possess normal modes, but they
tend to be localized around wave attractors;
\citealt{1997JFM...341...77R}.)

There do exist special, purely toroidal modes, for which
\begin{equation}
  \omega=-\f{2\Omega}{m+1},
\end{equation}
\begin{equation}
  c_m=r^{m-1},
\end{equation}
\begin{equation}
  W_{m+1}=-2m\Omega\tilde q_{m+1}r^{m+1},
\end{equation}
while all other components vanish
\citep{1997JFM...341...77R}.  These solutions exist
independent of the radial extent of the annulus.  Note that there is
only one such mode for each azimuthal wavenumber~$m$.

Normal modes involving some radial motion can be sought by expressing
them in the basis of exponential solutions described above.  Setting
the determinant of the matrix of coefficients to zero yields (in the
case $m=2$) a polynomial equation of degree $L(L-2)/2$ in
$\omega/\Omega$.  Numerical investigation suggests that the roots are
all real and lie in the interval $(-2,2)$.  These solutions presumably
give rise to a complete set of normal modes for the truncated problem.
However, the frequencies and eigenfunctions of the modes do not appear
to converge as $L$ is increased, which suggests that the spherical
annulus does not possess normal modes other than the purely toroidal
ones.

In the special case of a coreless planet ($\alpha=0$) normal modes do
exist \citep{1968Greenspan}.  The simplest symmetric modes for $m=2$,
for example, are
\begin{equation}
  a_n=r(r^2-R^2)\delta_{n2},
\end{equation}
\begin{equation}
  \f{\omega}{\Omega}=-\f{1}{2}\left(1\pm\sqrt{\f{15}{7}}\right)=-1.2319,\,0.2319.
\end{equation}
It might be though that such modes could be excited resonantly by
tidal forcing.  However, the exact solution of the radially
forced problem for $\ell=m=2$ in a coreless planet is
\begin{equation}
  a_n=\f{r}{R}\,\delta_{n2}
\end{equation}
and does not allow for any such resonances.  (In cylindrical polar
coordinates $(s,\phi,z)$ this solution is $u_s=s/R$, $u_\phi=\rmi
s/R$, $u_z=0$.)  This is also a solution of the nonlinear
problem, because $\bu\cdot\bnabla\bu={\bf0}$.  In a similar way, the
Roche--Riemann ellipsoids provide the nonlinear solutions of certain
problems of tidal forcing in the case of a homogeneous incompressible
planet without a core \citep{2008arXiv0812.1028G}.

\section{Numerical analysis}
\label{s:numerical}

\begin{figure*}
\centerline{\epsfysize=4.3cm\epsfbox{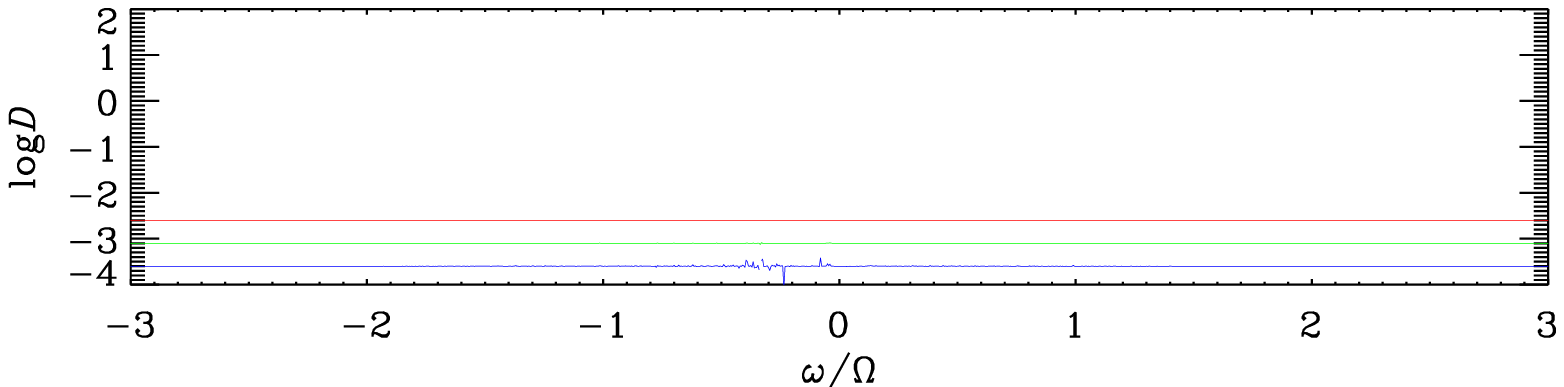}}
\centerline{\epsfysize=4.3cm\epsfbox{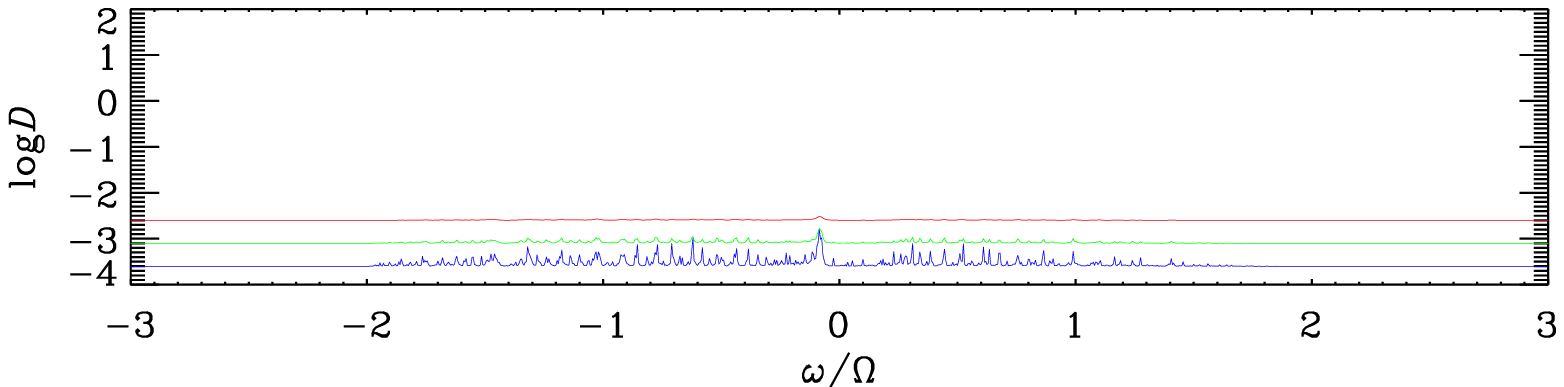}}
\centerline{\epsfysize=4.3cm\epsfbox{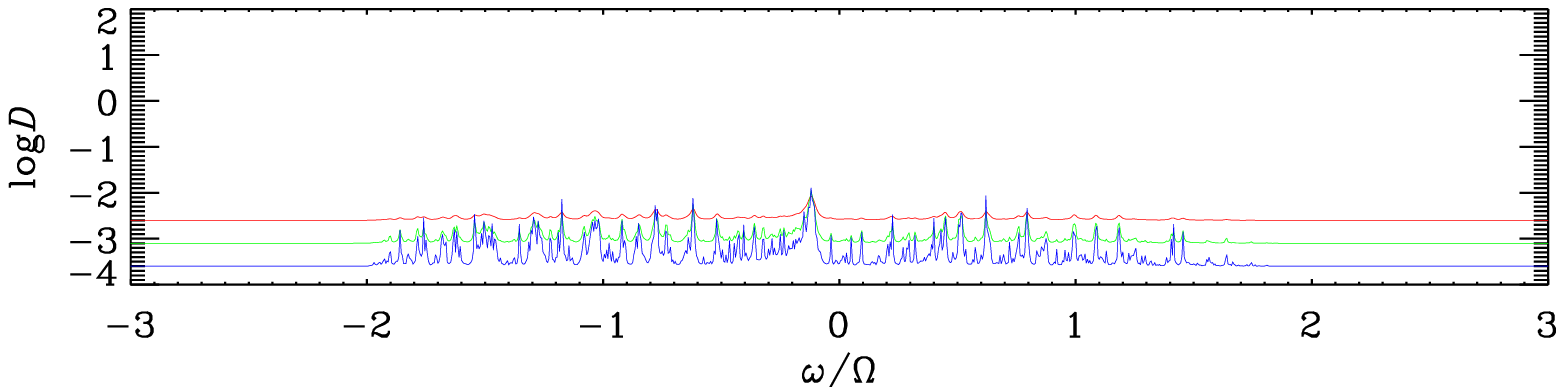}}
\centerline{\epsfysize=4.3cm\epsfbox{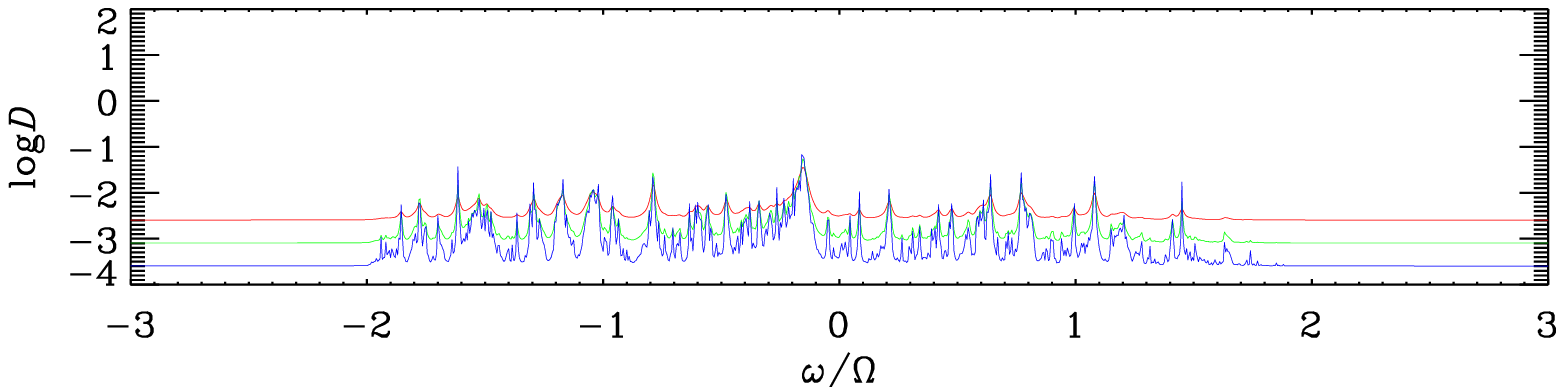}}
\centerline{\epsfysize=4.3cm\epsfbox{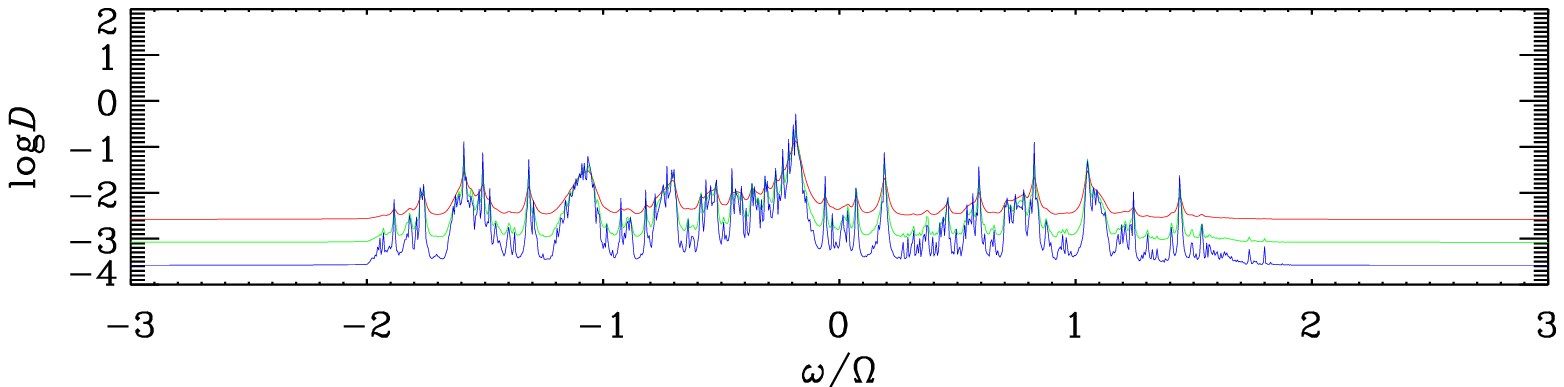}}
\caption{Variation of the total dissipation rate (in units of $\rho
R^3U^2\Omega$) with tidal frequency, for a fluid with a frictional
force.  The radial velocity is forced at the outer boundary in the
form of an $\ell=m=2$ spherical harmonic of amplitude $U$.  The
fractional core radius is $\alpha=0.1$, $0.2$, $0.3$, $0.4$ and $0.5$
(panels from top to bottom).  The frictional damping coefficient
$\gamma$ is given by $\gamma/\Omega=10^{-2}$, $10^{-2.5}$ and
$10^{-3}$ (red, green and blue curves, from top to bottom for the most
part).  In all figures, `log' denotes a base-10 logarithm.}
\label{f:fricover1}
\end{figure*}

\begin{figure*}
\centerline{\epsfysize=4.3cm\epsfbox{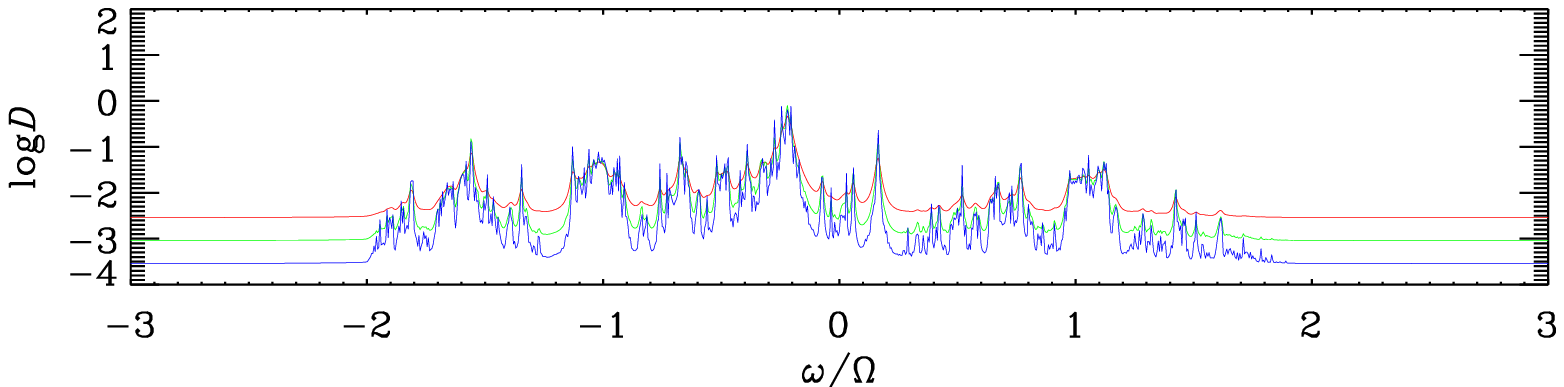}}
\centerline{\epsfysize=4.3cm\epsfbox{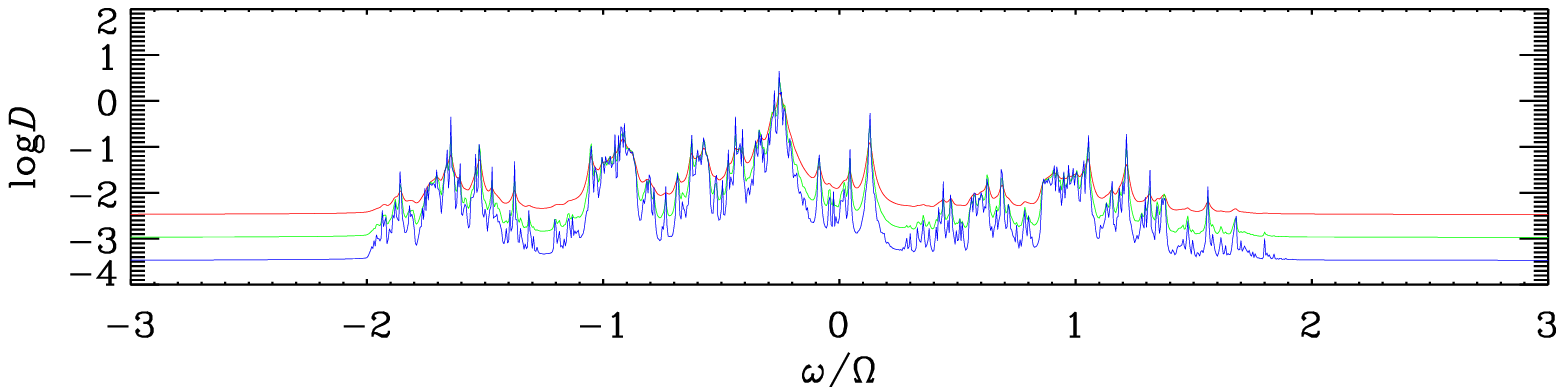}}
\centerline{\epsfysize=4.3cm\epsfbox{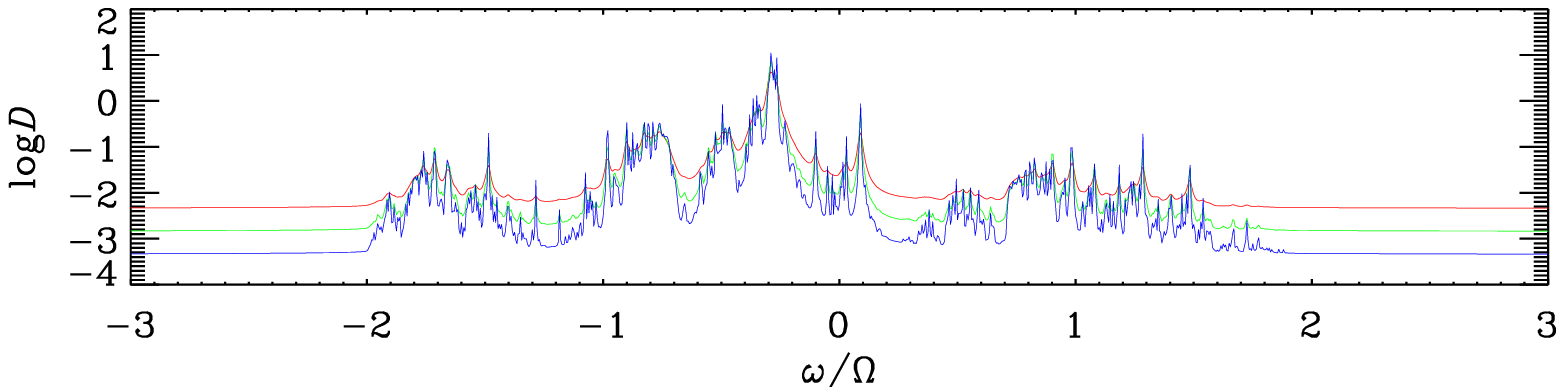}}
\centerline{\epsfysize=4.3cm\epsfbox{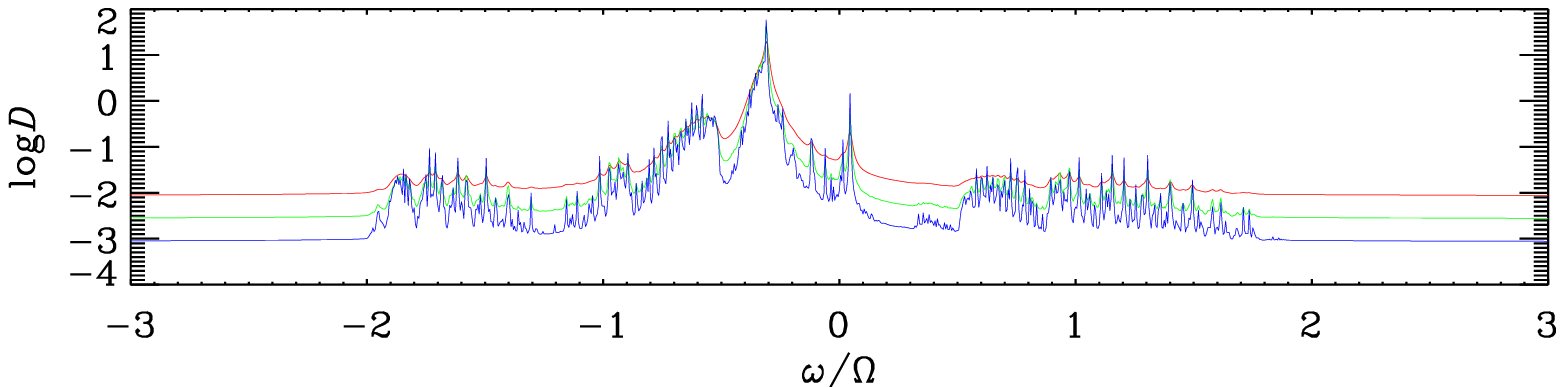}}
\centerline{\epsfysize=4.3cm\epsfbox{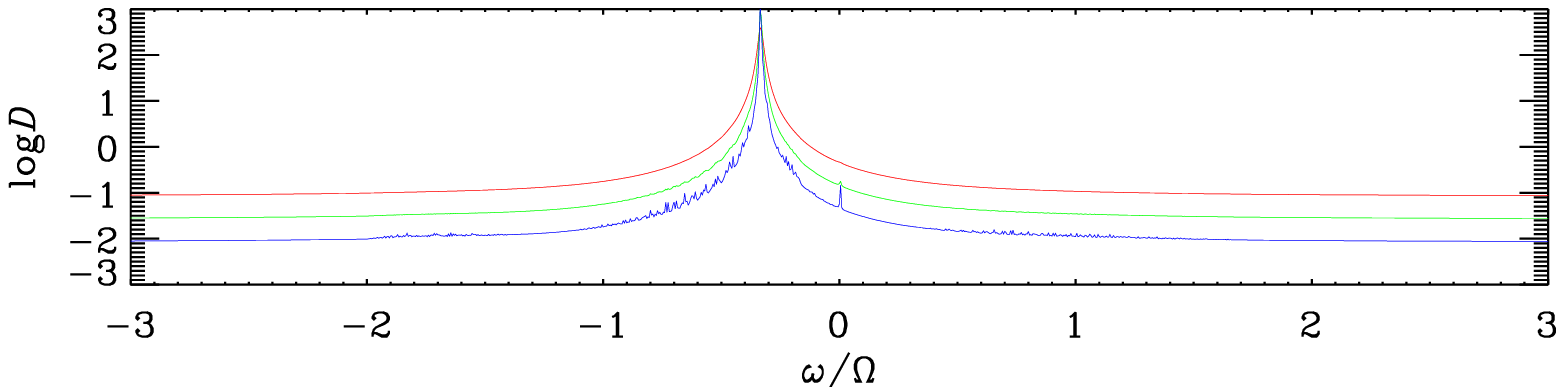}}
\caption{Continuation of Fig.~\ref{f:fricover1} for fractional core radii $\alpha=0.6$, $0.7$, $0.8$, $0.9$ and $0.99$ (panels from top to bottom).  Note the different vertical scale in the last case.}
\label{f:fricover2}
\end{figure*}

\begin{figure*}
\centerline{\epsfbox{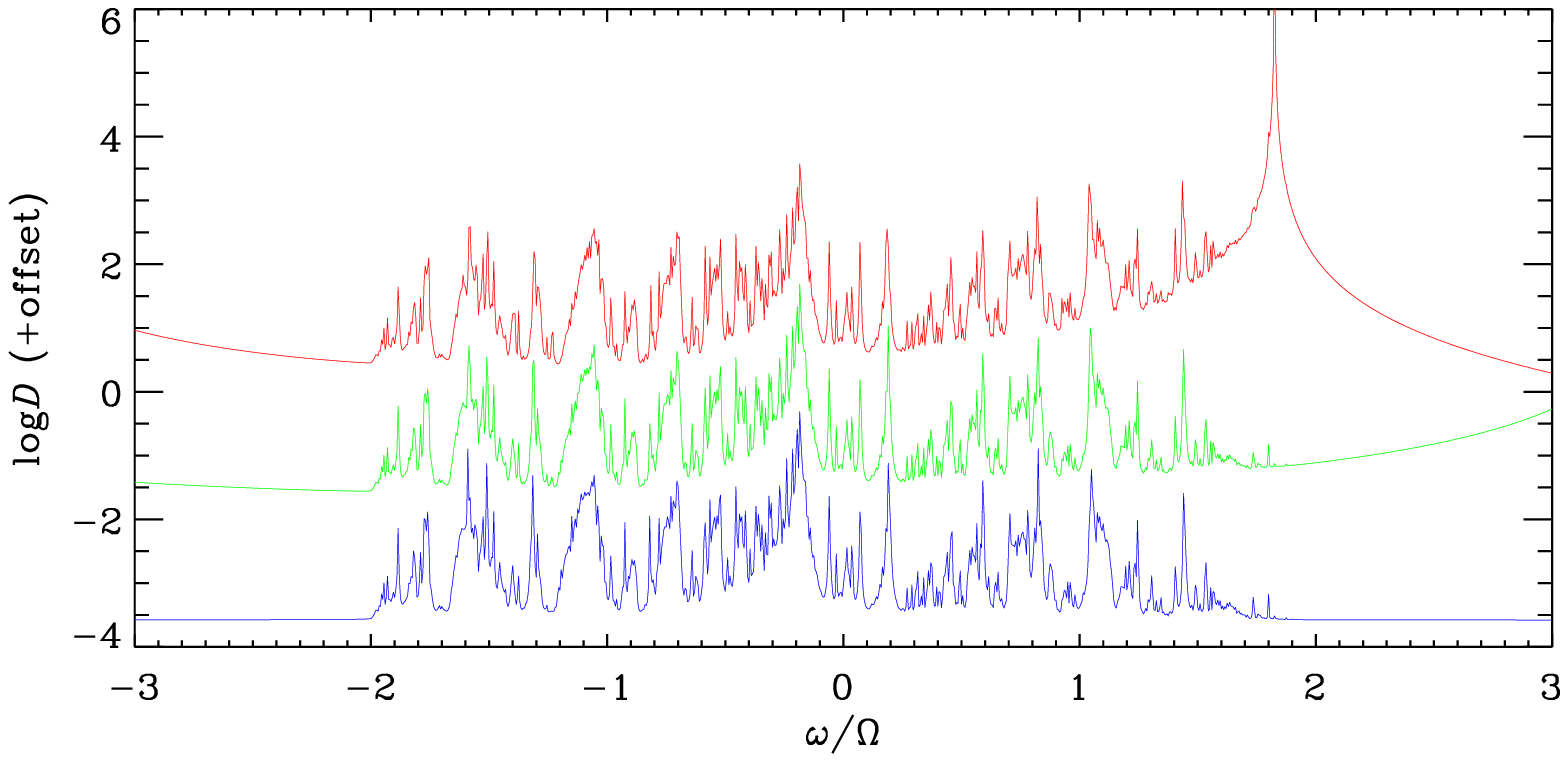}}
\caption{Comparison of tidal and radially forced problems.  The bottom
(blue) curve shows the total dissipation rate (in units of $\rho
R^3U^2\Omega$) for $\alpha=0.5$ and $\gamma/\Omega=10^{-3}$, as in the
bottom panel of Fig.~\ref{f:fricover1}.  The middle (green) and top
(red) curves show the dissipation rate for the tidally forced problem
in a planet with dynamical frequency $(g/R)^{1/2}=5\,\Omega$ and
$3\,\Omega$, respectively, and with mean density equal to the fluid
density.  The latter curves are vertically offset by 2 and 4 units,
respectively, for clarity.  The comparison is made by equating the
forced radial velocity amplitude $U$ with $\omega(5/2)(\Psi/g)$.}
\label{f:tidal}
\end{figure*}

\subsection{Method of solution}
\label{s:method}

We focus on the $\ell=m=2$ type of quadrupolar forcing that is of
greatest importance in applications.  For the viscous problem, we
solve equations~(\ref{aneq})--(\ref{dneq}) and the appropriate
boundary conditions by the pseudospectral method as described in
\citet{2004ApJ...610..477O}.  This makes use of Chebyshev collocation
on a grid of $N+1$ points in $r$ (the Gauss--Lobatto nodes).  The
system of equations is also truncated at spherical harmonic degree
$L$.  The problem involves a large block-tridiagonal matrix and the
solution is obtained by a standard direct method.  The total
dissipation rate is computed by two independent methods: from the
direct dissipation integral (equation~\ref{dint}) using a Chebyshev
quadrature formula, and from the rate of working at the surface
(equation~\ref{dsurf1} or~\ref{dsurf2} as appropriate).

For the frictional problem, we apply Chebyshev collocation to
equation~(\ref{aneq2}) and the appropriate boundary conditions.  This
has the advantage that the blocks of the tridiagonal matrix are
smaller because the other variables have been eliminated in favour of
the radial velocity, and therefore a higher resolution can be reached.
(In this case the Chebyshev variable is taken to be linearly related
to $y$ rather than $r$.)

To validate these methods, we checked that the solutions found by the
exponential technique described in Section~\ref{frictional}, which
works very well for modest values of $L$, agree with those obtained by
the pseudospectral method.  We verified numerically the analytical
expressions obtained in Sections~\ref{fricnr} and~\ref{viscnr} for the
dissipation rate in the non-rotating tidal problem.  We also confirmed
that the dissipation rates calculated by the integral and surface
formulae agree well in all cases.

\subsection{Overview of results}

In Figs~\ref{f:fricover1} and~\ref{f:fricover2} we present an overview
of the behaviour of the dissipation rate with tidal frequency in the
radially forced problem with a frictional force, for different values
of the fractional core radius $\alpha$ and the frictional damping
coefficient $\gamma$.  These results were obtained with numerical
resolutions of $L=N=200$ or~$300$, which is adequate for this kind of
overview.  A higher resolution is generally required for smaller
values of $\gamma$.

Outside the range $-2<\omega/\Omega<2$ the dissipation rate is
essentially independent of $\omega$ and proportional to $\gamma$.  No
inertial waves are excited and the response is smooth and
uninteresting.  The dissipation rate is greater in a thin shell
because of the larger value of the parameter $B$.

Within the range $-2<\omega/\Omega<2$ the dissipation rate is enhanced
through the excitation of inertial waves, as found by
\citet{2004ApJ...610..477O}.  There is a complicated dependence on
frequency.  In the frequency intervals in which the dissipation rate
is most enhanced, it appears to be roughly independent of the
frictional damping coefficient.

This enhancement of the dissipation rate increases systematically with
the size of the core.  In the case of a very thin shell
($\alpha=0.99$), we observe instead a strong resonance with a
Rossby wave or r~mode at $\omega=-\Omega/3$, as described below.

\subsection{Resonance with a Rossby wave or r~mode in a thin shell}

The limit of a thin shell, $1-\alpha=\epsilon\ll1$, is of some
interest and can be related to problems studied by Laplace
\citep[see][and references therein]{1932Lamb}.  In the absence of
viscosity, motions are possible in which, to a first approximation,
$a_n$ varies linearly with $r$, vanishing at the inner radius, and is
much smaller than $rb_n$ or $rc_n$.  Under these circumstances the
`traditional approximation' and Laplace's tidal equations are
applicable.  In this limit the shell possesses free modes of
oscillation which depend on the dimensionless parameter
$4\Omega^2R^2/gh$, where $h=\epsilon R$ is the depth of the ocean
\citep{1968RSPTA.262..511L}.  When this parameter is much less than
unity, simple analytical solutions emerge with frequency
$\omega=-2m\Omega/n(n+1)$ for integers $n\ge m$.  These are the
planetary or Rossby waves, which are well separated from surface
gravity waves in the limit under consideration.  Related solutions can
be identified in stably stratified stellar atmospheres, where they are
known as r~modes \citep{1978MNRAS.182..423P}.

The resonance described above occurs with the $(m=2,n=3)$
Rossby wave or r~mode at $\omega=-\Omega/3$.  (Note that this is not
equivalent to the special, purely toroidal mode discussed in
Section~\ref{s:modes}, which has a frequency of $\omega=-2\Omega/3$
and involves no radial motion.)  The appropriate balance in
equations~(\ref{cneq}) and~(\ref{dneq}) is then
\begin{equation}
  -\rmi\omega_nc_n-2\Omega[(n-1)\tilde q_nb_{n-1}+(n+2)\tilde q_{n+1}b_{n+1}]\approx0,
\end{equation}
\begin{equation}
  \f{\rmd a_n}{\rmd r}-n(n+1)b_n\approx0.
\end{equation}
The boundary conditions determine that $\rmd a_n/\rmd r=(U/\epsilon
R)\delta_{n\ell}$ and we easily obtain $b_n$ and $c_n$ from the above
equations.  In the case $\ell=m=2$ investigated here, the toroidal
velocity component $c_3$ is excited and becomes large when $\omega_3$
is small, which occurs close to $\omega/\Omega=-1/3$.  The resulting
dissipation rate is
\begin{equation}
  D\approx\f{\rho\gamma R^3U^2}{12\epsilon}\left(1+\f{32\Omega^2}{63|\omega_3|^2}\right).
\end{equation}
This expression provides an excellent fit to the numerically
determined dissipation rate in the frictional problem for
$\alpha=0.99$.  Note that $D$ is inversely proportional to the
thickness of the shell.  This result occurs because the fast
barotropic flow (i.e.~horizontal motion independent of depth) excited
in a thin shell is efficiently damped by the artificial frictional
force in this model; a different behaviour would occur in a viscous
fluid.  Nevertheless, it is well known from the case of the Earth that
highly efficient tidal dissipation can occur in a shallow ocean.  The
case of the Earth's ocean is of course enormously complicated by its
irregular shape and depth.  Mode conversion and turbulence provide
channels for dissipation.

\subsection{Comparison of tidal and radially forced problems}

The original tidal problem, described in Section~\ref{s:tidal}, gives
results very similar to those of the simplified, radially forced
problem.  Some additional parameters are required to set up the tidal
problem: the ratio of the dynamical frequency $(g/R)^{1/2}$ to the
spin frequency $\Omega$, and the ratio $\beta$ of the fluid density to
the mean density of the planet.  In Fig.~\ref{f:tidal} we compare the
dissipation rates in the tidal and radially forced problems, with
$\alpha=0.5$, $\beta=1$ and $\gamma/\Omega=10^{-3}$.  In order to make
the comparison we assume that the radial velocity at the surface is
determined by equation~(\ref{tidalbc}) in the low-frequency limit in
which the $W_n$ term (and also the viscous term) is neglected.  This
is equivalent to saying that the radial displacement of the surface is
$-(5/2)\Psi/g$, where the factor of $5/2$ comes from the self-gravity
of the fluid.

When the planet is slowly rotating, in the sense that
$\Omega\ll(g/R)^{1/2}$, the agreement is excellent and shows that the
excitation of inertial waves is not affected by the freedom of the
outer boundary.  For more rapidly rotating planets the influence of
surface gravity waves becomes apparent in this range of frequencies.
For such rapidly rotating planets the centrifugal distortion of the
fluid ought to be taken into account.

\subsection{Investigation of a restricted frequency interval}

\begin{figure}
\centerline{\epsfysize8cm\epsfbox{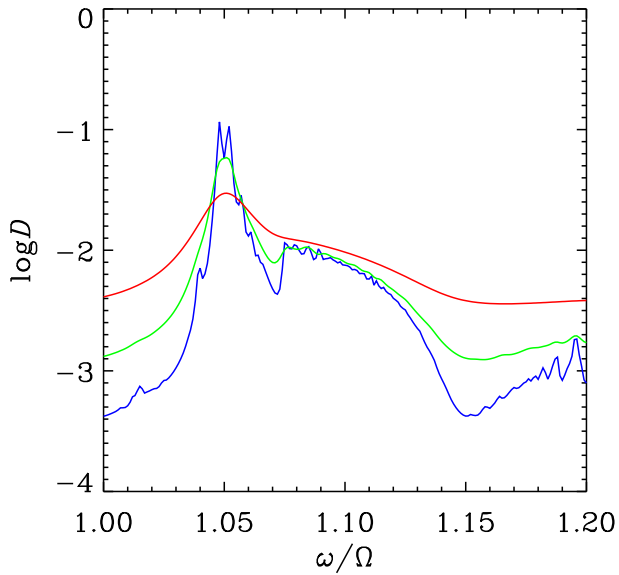}}
\caption{Expanded view of the bottom panel of Fig.~\ref{f:fricover1},
i.e.\ for $\alpha=0.5$ and $\gamma/\Omega=10^{-2}$, $10^{-2.5}$ and
$10^{-3}$.}
\label{f:fric}
\end{figure}

In order to investigate the response in greater detail, we restrict
our attention to the case of a fractional core radius $\alpha=0.5$ and
to a certain range of frequencies, $1<\omega/\Omega<1.2$.  This
interval is chosen partly because, as described below, it contains two
relatively simple wave attractors.  Fig.~\ref{f:fric} shows an
expanded view of the bottom panel of Fig.~\ref{f:fricover1}, using a
higher frequency resolution.  A higher numerical resolution (up to
$L=N=800$) is also used in order to ensure that the results are
adequately converged for the appearance of this figure.

\begin{figure}
\centerline{\epsfysize8cm\epsfbox{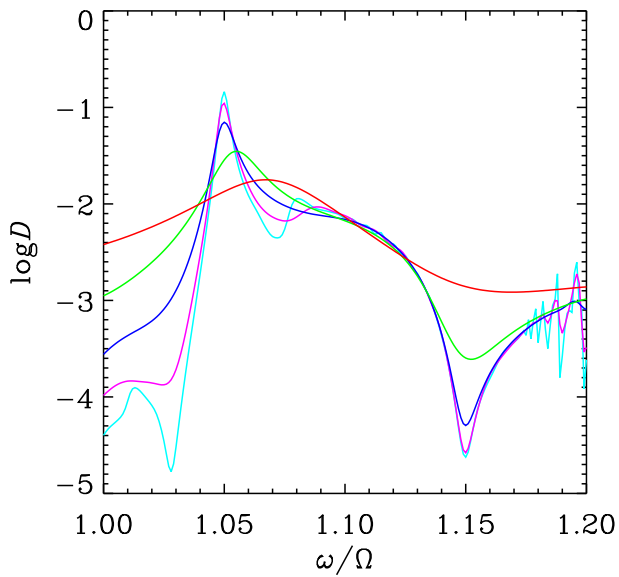}}
\caption{Equivalent of Fig.~\ref{f:fric} for a viscous fluid with $\nu/R^2\Omega=10^{-4}$, $10^{-5}$, $10^{-6}$, $10^{-7}$ and $10^{-8}$ (red, green, blue, magenta and cyan curves).}
\label{f:visc}
\end{figure}

Fig.~\ref{f:visc} shows equivalent results obtained for a viscous
fluid.  In this case a smaller numerical resolution (up to $L=N=400$)
is achieved and the results are adequately converged except possibly
in some cases for the lowest viscosities.

\begin{figure}
\centerline{\epsfysize8cm\epsfbox{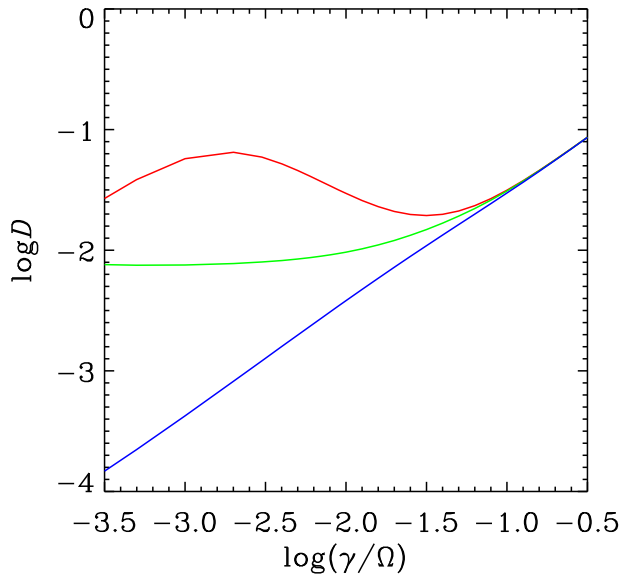}}
\caption{Frictional dissipation versus frictional damping coefficient for $\alpha=0.5$ and $\omega/\Omega=1.05$, $1.10$ and $1.15$ (red, green and blue curves, from top to bottom).}
\label{f:varygamma}
\end{figure}

\begin{figure}
\centerline{\epsfysize8cm\epsfbox{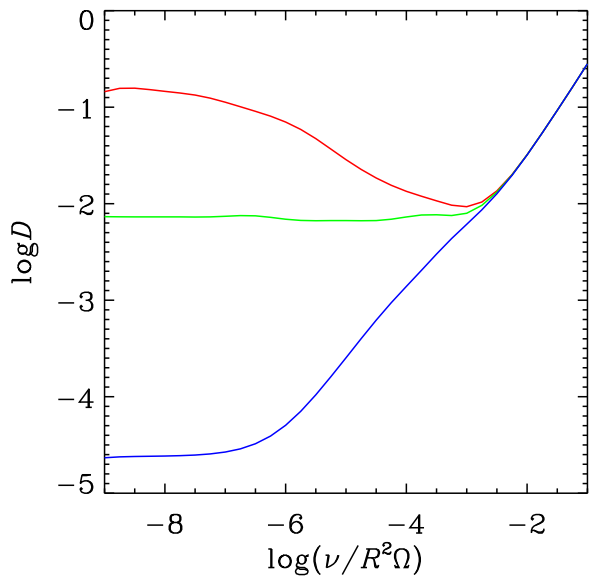}}
\caption{Viscous dissipation versus viscosity for $\alpha=0.5$ and $\omega/\Omega=1.05$, $1.10$ and $1.15$ (red, green and blue curves, from top to bottom).}
\label{f:varynu}
\end{figure}

We also show the variation of the dissipation rate with the
coefficient of friction or viscosity for three frequencies in this
interval, $\omega/\Omega=1.05$, $1.10$ and $1.15$, in
Figs~\ref{f:varygamma} and~\ref{f:varynu}.  To obtain adequately
converged results for the lowest values of $\gamma$ or $\nu$, a very
high resolution (up to $L=N=1600$) is required.

Perhaps the simplest behaviour occurs close to $\omega/\Omega=1.10$.
Here the dissipation rate converges to a non-negligible value as
either $\gamma\to0$ or $\nu\to0$.  The same limiting value is obtained
in the frictional and viscous problems and depends smoothly on
$\omega$ within a certain interval.  This behaviour is closely
reminiscent of that found in the analysis of a simple wave attractor
\citep{2005JFM...543...19O}.  We will see below that two distinct
types of wave attractor are active in this range of frequencies.

The maximum dissipation rate occurs close to $\omega/\Omega=1.05$.  In
the frictional problem $D$ does not appear to converge as
$\gamma\to0$, although the ultimate behaviour is not clear from the
limited range of parameters that can be investigated numerically.  In
the viscous problem the dissipation rate gives the appearance of
convergence as $\nu\to0$ but it may decrease below
$\nu/R^2\Omega=10^{-9}$.

A much lower dissipation rate is found close to $\omega/\Omega=1.15$.
Here $D$ appears to converge to a very small value as $\nu\to0$.  It
may do so in the frictional problem at lower values of $\gamma$ than
are achieved here.

\subsection{Spatial structure of the response}

\begin{figure}
\centerline{\epsfysize8cm\epsfbox{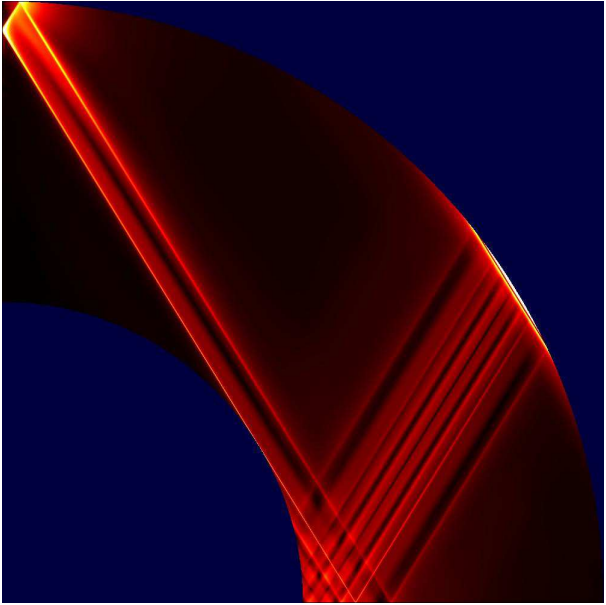}}
\caption{Structure of the velocity perturbation $|\bu|$ in a
meridional quarter-plane for the case $\alpha=0.5$, $\omega/\Omega=1.05$,
$\gamma/\Omega=10^{-3}$, $L=N=1600$.  A linear colour table is used.}
\label{f:image1.05}
\end{figure}

\begin{figure}
\centerline{\epsfysize8cm\epsfbox{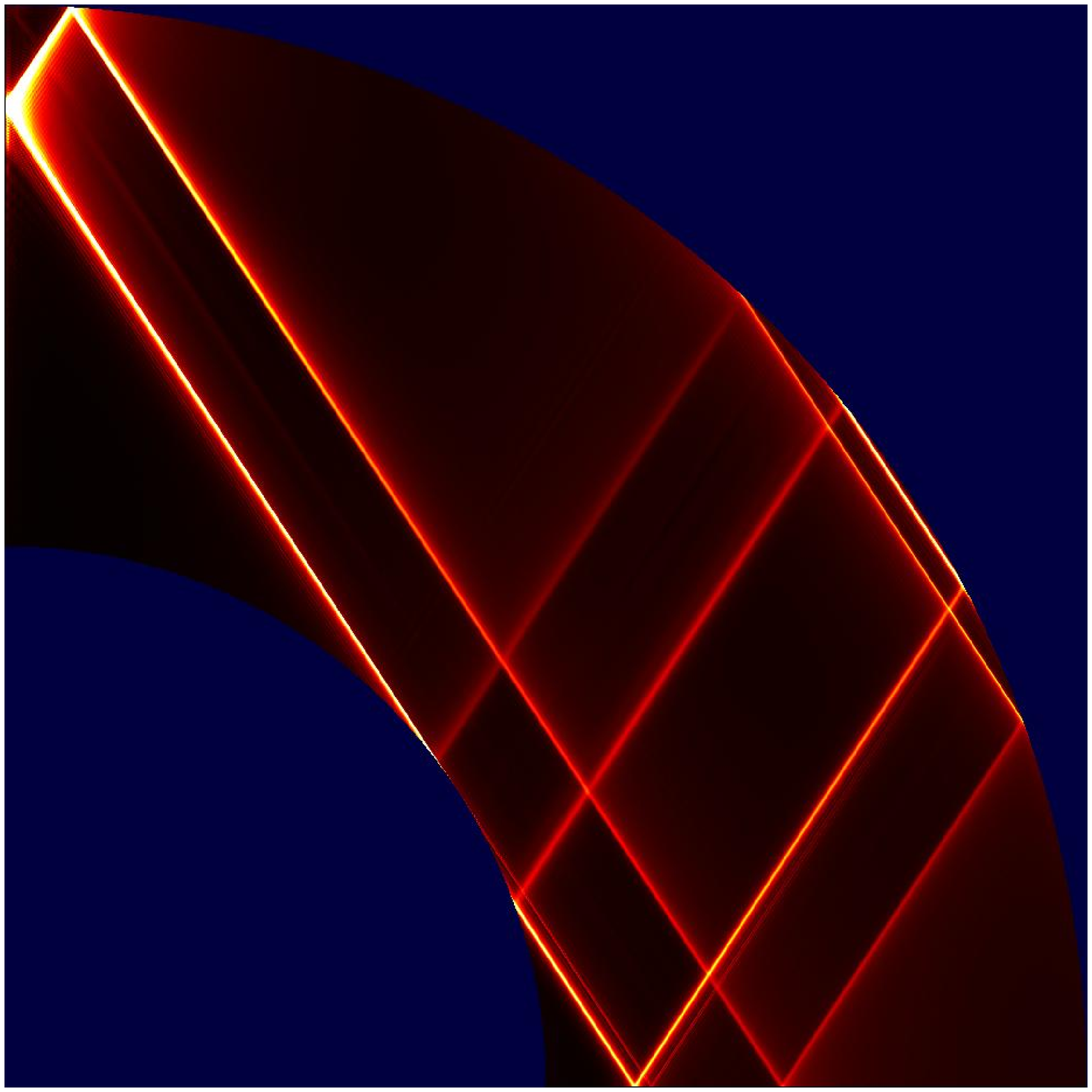}}
\caption{As Fig.~\ref{f:image1.05} but for $\omega/\Omega=1.10$.}
\label{f:image1.10}
\end{figure}

\begin{figure}
\centerline{\epsfysize8cm\epsfbox{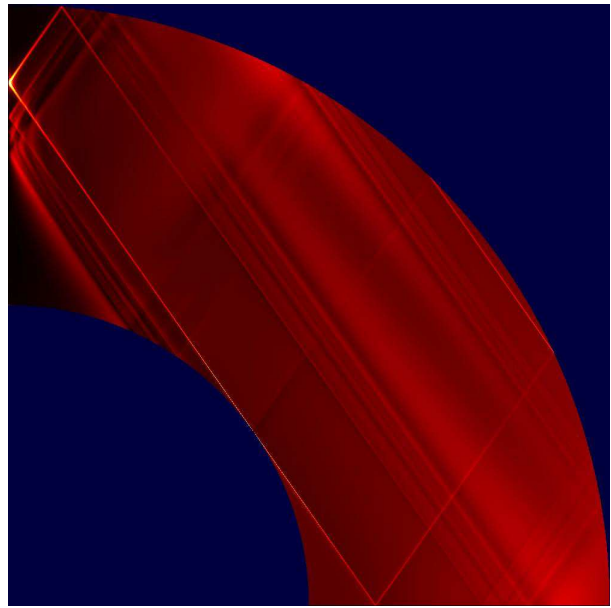}}
\caption{As Fig.~\ref{f:image1.05} but for $\omega/\Omega=1.15$.}
\label{f:image1.15}
\end{figure}

We now examine the spatial structure of the response in the meridional
plane.  Figs~\ref{f:image1.05}--\ref{f:image1.15} show the magnitude
of the velocity perturbation, $|\bu|$, for frequencies
$\omega/\Omega=1.05$, $1.10$ and $1.15$, in the frictional problem
with $\gamma/\Omega=10^{-3}$.  In each case the disturbance is
concentrated near selected inertial-wave rays, which are
characteristics of the Poincar\'e equation.

\begin{figure}
\centerline{\epsfysize8cm\epsfbox{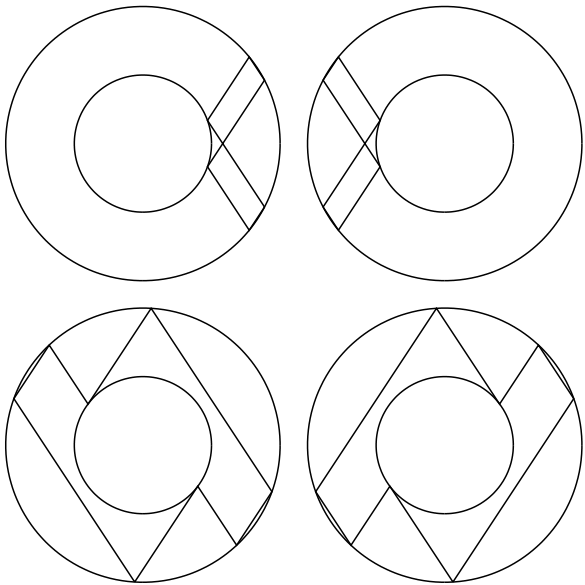}}
\caption{Wave attractors in a spherical annulus with $\alpha=0.5$ at frequency $\omega/\Omega=1.10$.}
\label{f:attractors}
\end{figure}

Relatively simple wave attractors exist in this range of frequencies.
The symmetrical pair of attractors involving four reflections on the
outer boundary, illustrated in the upper part of
Fig.~\ref{f:attractors}, exists for $1.076<\omega/\Omega<1.148$; a
second pair of attractors, involving six outer reflections, exists for
$1.091<\omega/\Omega<1.165$.  The focusing power of an attractor
depends on frequency in a systematic way within the bandwidth of the
attractor, as its shape is adapted \citep{2001JFM...435..103R}.  The
first one is strongest at the right-hand end of its bandwidth, while
the second is strongest at the left-hand end.

The response at frequency $1.10$ (Fig.~\ref{f:image1.10}) shows that
the two types of attractor that exist at this frequency are both
activated.  The wave energy is concentrated around the attractors in a
region whose width is proportional to $\gamma$.  This behaviour, and
the convergence of the dissipation rate as $\gamma\to0$ or $\nu\to0$,
are entirely analogous to that found by \citet{2005JFM...543...19O}.

Also seen in this case, however, is a ray emerging from the critical
latitude on the inner boundary.  This ray is indeed the dominant
feature apparent at frequency $1.15$ (Fig.~\ref{f:image1.15}) where
the dissipation is relatively weak.  The attractor with six outer
reflections is present in the solution at frequency $1.15$ but is
weakly focusing and is not powerfully activated.  The dissipation rate
in the viscous problem appears to converge to a very low value as
$\nu\to0$ at this frequency.

The behaviour at frequency $1.05$ (Fig.~\ref{f:image1.05}), where the
dissipation is strongest, is harder to explain.  The ray dynamics in
the vicinity of this frequency is very complicated.  Although
extremely long attractors may exist (e.g.\ one at frequency $1.05$
involving 584 outer reflections) they are of no practical
significance.  However, there is a tendency for rays to be
concentrated temporarily into the region between the equator of the
inner boundary and the critical latitude on the outer boundary, where
the wave energy is seen to be prominent.  The rays escape from this
region and reenter it repeatedly.  It is difficult to identify
features of the ray dynamics that lead to a preference for the
frequency $1.05$.

The prominence of rays emerging from the critical latitude on the
inner boundary has been noticed before
\citep[e.g.][]{1999PhRvE..59.1789T,2004ApJ...610..477O}.

\subsection{Frequency-averaged dissipation rate}

\begin{figure}
\centerline{\epsfysize8cm\epsfbox{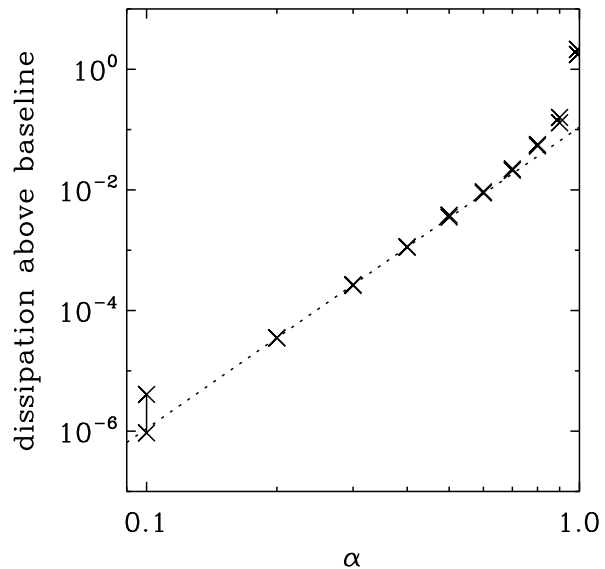}}
\caption{Estimates of the frequency-averaged dissipation rate over the interval $-2<\omega/\Omega<2$, minus the `baseline' dissipation outside this range, versus fractional core radius, for the frictional problem with $10^{-3}\le\gamma/\Omega\le10^{-2}$.  The dotted line represents $0.11\alpha^5$.}
\label{f:average}
\end{figure}

There is a strong systematic dependence of the dissipation rate on the
size of the core.  This effect is examined in Fig.~\ref{f:average},
where we plot an average dissipation rate versus $\alpha$.  The
average is obtained from the data for Figs~\ref{f:fricover1}
and~\ref{f:fricover2}, by taking the arithmetic mean of the
dissipation rates evaluated at 800 equally spaced frequencies in the
range $-2<\omega/\Omega<2$.  The `baseline' dissipation rate, which
occurs for frequencies outside this range, is subtracted from the
average, in order to isolate the contribution due to waves.  There may
be a systematic error in this procedure owing to the inadequate
sampling of narrow peaks in the curve.

The resulting values are approximately independent of $\gamma$ and
increase approximately as the fifth power of the core radius.
Interestingly, this dependence is predicted by the recent analysis of
\citet{2008arXiv0812.1028G}, who consider the scattering of the
equilibrium tide off the core, although they do not predict the
intricate variation with frequency.  A departure from this behaviour
occurs as $\alpha$ approaches~$1$, owing to the toroidal-mode
resonance in a thin shell.

\section{Discussion}
\label{s:discussion}

\subsection{Interpretation of enhanced dissipation}

The dynamics of inertial waves in a spherical annulus is very rich and
a great deal remains to be explained.  In earlier work
\citep{2005JFM...543...19O} we considered a simpler type of domain in
which all wave energy is focused towards a single wave attractor and a
dissipation rate is obtained that is asymptotically independent of the
viscosity or other small-scale damping mechanism for the waves.  In a
spherical annulus, the closest parallel to this behaviour occurs in
situations such as that illustrated in Fig.~\ref{f:image1.10},
although in fact more than one attractor is activated there.  The
spherical annulus is more complicated because simple attractors occupy
only a limited range of frequencies, and also because of the existence
of a critical latitude at which rays are tangent to the inner
boundary.  In fact these features are related to each other, because
the range of frequencies in which an attractor exists is bounded by
situations in which its rays approach the critical latitude on the
inner and outer boundaries \citep{2001JFM...435..103R}.

\begin{figure}
\centerline{\epsfysize7cm\epsfbox{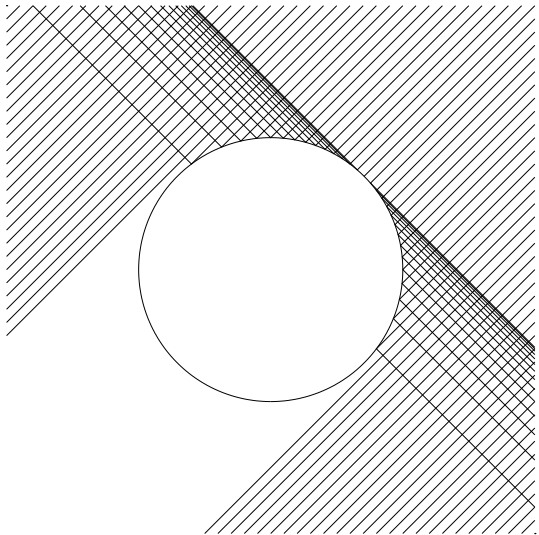}}
\medskip
\caption{Formation of a singularity at the critical latitude through the reflection of rays (equally spaced lines incident from the top right) from the inner boundary.}
\label{f:critlat}
\end{figure}

Both attractors and critical latitudes concentrate wave energy and
promote enhanced dissipation.  A smooth beam of waves reflecting from
the inner boundary generates a singularity in an inviscid fluid, as
illustrated in Fig.~\ref{f:critlat}.  This effect depends smoothly on
frequency in the range $-2<\omega/\Omega<2$.  When the singular beam
emanating from the critical latitude is resolved by viscosity,
enhanced dissipation occurs but this alone does not account for the
behaviour seen close to $\omega/\Omega=1.05$ in Fig.~\ref{f:visc}; it
is more suitable as an explanation of the behaviour close to
$\omega/\Omega=1.00$, where $D$ scales approximately with $\nu^{1/2}$.

The behaviour of rays alone is not sufficient to explain the shape of
the dissipation curve.  This can be seen from the fact that $D$
depends on the sign of $\omega/\Omega$ and on the value of $m$ (not
shown here), whereas the rays are insensitive to these properties.
Even in the case of a simple wave attractor, the full calculation of
the asymptotic dissipation rate depends on a detailed analysis of the
forcing accumulated around each ray circuit
\citep{2005JFM...543...19O}.

Some time after the calculations described in this paper were
completed, we received a preprint from \citet{2008arXiv0812.1028G}.
They treat the equilibrium tide as a large-scale inertial wave that
reflects from the inner boundary.  The intense beam emanating from the
neighbourhood of the critical latitude produces enhanced dissipation,
occurring in their analysis through nonlinearity.  This theory has an
attractive simplicity to it and appears to predict the correct scaling
of the frequency-averaged dissipation rate on the size of the core.
It does not predict a complicated frequency-dependence because further
reflections are not considered.  Indeed, as is mentioned below, the
various physical circumstances in planets and stars may not always
permit the multiple reflections that occur in the simple model
considered in the present paper.

After this paper was submitted, we also received a manuscript
from M.~Rieutord and L.~Valdettaro, who have considered a similar
problem and have been able to relate some aspects of the
frequency-dependence of the dissipation rate to the properties of wave
attractors, viscous normal modes and critical latitudes.

\subsection{Astrophysical and planetary applications}

The dissipation rates $D$ (expressed in units of $\rho R^3U^2\Omega$)
plotted in various figures in this paper can be converted into
modified tidal quality factors $Q'$ according to the relation
\[
  \f{1}{Q'}=\f{4}{5}\beta(1-{\textstyle\f{3}{5}}\beta)^{-2}\f{\Omega|\omega|}{g/R}\left(\f{D}{\rho
  R^3U^2\Omega}\right).
\]
For a fixed value of $\omega/\Omega$ in the range of inertial waves,
the efficiency of tidal dissipation through this mechanism therefore
scales with the square of the ratio of the spin frequency to the
dynamical frequency of the body \citep{2004ApJ...610..477O}.  This
fact should always be borne in mind when comparing solar system
planets with hot Jupiters and hot Neptunes, which are expected to
rotate more slowly as a result of tidal evolution.  The wide variation
in $D$ indicates that a very broad range of $Q'$ values can occur
through this mechanism.  The strong frequency-dependence of $Q'$ can
have important consequences for tidal evolution, for example in
shaping the satellite systems of the giant planets.

We note that the tidal quality factor provides only a
conventional and, for some purposes, convenient parametrization of the
efficiency of tidal dissipation, which can involve complicated
processes of wave excitation, propagation and damping.  It depends as
much on the capacity of the system to respond to oscillatory forcing
as on its ability to dissipate the response.

\subsection{Topics for further investigation}

The problem studied in this paper is intentionally simplified in order
to try to isolate an already complicated physical process.  We mention
here some of the topics that require further investigation.

The quality of wave reflection from the boundaries of the fluid region
supporting inertial waves should be analysed.  Depending on the type
of planet or star being considered, these boundaries may be free
surfaces, interfaces with solid or denser fluid material, or borders
between convective and radiative zones.  Internal reflection of
inertial waves from discontinuities such as a putative first-order
phase transition in giant planets are also potentially of importance.

The propagation of inertial waves could be affected by differential
rotation, magnetic fields, buoyancy effects and the centrifugal
distortion of the body.  It would also be of interest to study an
initial-value problem rather than assuming a response that is harmonic
in time.

Dissipation mechanisms for inertial waves at small scales include
viscous damping, interaction with turbulent convection, Ohmic
dissipation through coupling with magnetic fields, and nonlinear
dissipation through wave breaking or parametric instability.  While
the analysis of a simple wave attractor indicated a total dissipation
rate that is asymptotically independent of the small-scale processes,
a more complicated picture arises from the present paper and it may be
necessary to understand the damping mechanisms in greater detail.

The dependence of the dissipation rate on the size of the core should
be investigated in more realistic models where the fluid is
compressible and the core is deformable.  The role of resonances with
global normal modes (if they exist) in compressible models needs to be
clarified.  The relevance of boundary layers on the surface of the
core, obviated here through the use of stress-free boundary
conditions, is worthy of consideration.  Dissipation within the core
itself is yet another area of uncertainty.

\subsection{Elliptical instability}

A quite different paradigm for tidal dissipation in rotating planets
and stars could emerge through a consideration of elliptical
instabilities \citep[][and references therein]{2002AnRFM..34...83K}.
If the linear tidal response is considered to provide merely an
elliptical distortion of the streamlines of the rotating fluid, as is
true in the case of a full sphere of incompressible fluid, then the
secondary instabilities of the elliptical flow can produce turbulence
and enhanced tidal dissipation.  Analytical, numerical and
experimental investigations have been motivated mainly by the case of
the Earth's core but also by Io \citep{1998GeoRL..25..603K} and by
accretion discs in binary stars
\citep{1993ApJ...406..596G,1993ApJ...419..758L,1994ApJ...422..269R}.
This nonlinear mechanism may be more relevant for tides in extrasolar
planets, while the much weaker tides in solar-system planets may
depend on linear mechanisms such as those described in this paper.
Since the linear response in a spherical annulus departs significantly
from a simple elliptical distortion, it would be of interest to
understand the nonlinear outcome in the presence of a core.

\section{Conclusion}
\label{s:conclusion}

In this paper we have studied the tidal forcing, propagation and
dissipation of linear inertial waves in a rotating fluid body.  The
intentionally simplified model involves a perfectly rigid core
surrounded by a deep ocean consisting of a homogeneous incompressible
fluid.  Centrifugal effects are neglected, but the Coriolis force is
considered in full, and dissipation occurs through viscous or
frictional forces.  We introduced a further simplification by
replacing the free outer surface with a boundary on which the radial
velocity is specified, which closely mimics the effects of tidal
forcing at low frequencies.  Various analytical results provide tests
of the numerical methods for calculating forced linear waves.

The dissipation rate exhibits a complicated dependence on the tidal
frequency and generally increases strongly with the size of the core.
In certain intervals of frequency, efficient dissipation is found to
occur even for very small values of the coefficient of viscosity or
friction, which is promising for astrophysical and planetary
applications.  In restricted intervals, the wave energy is focused
towards relatively simple wave attractors and a well defined
asymptotic dissipation rate is achieved.  However, a more typical
behaviour is that the inertial waves propagate in a very complicated
way around the fluid annulus.  The critical latitude on the inner
boundary plays an important role in the solutions.  While the
dissipation rate is enhanced in a strongly frequency-dependent manner,
it may not converge in the limit of small viscosity in the same way as
for a wave attractor.  In the limit of a thin fluid shell, the
dissipation rate can be greatly enhanced through a traditional type of
resonance with a toroidal mode (r~mode or Rossby wave).

The model adopted in this paper is deliberately oversimplified,
although it may be more or less applicable to planets or moons
involving a deep ocean.  We have pointed out numerous avenues for
further investigation, including aspects of internal wave propagation,
reflection and dissipation, and nonlinear behaviour such as the
elliptical instability.

\section*{acknoweldgments}

I thank Jeremy Goodman and the referee, Leo Maas, for helpful
comments.

\label{lastpage}

\end{document}